# Live Streaming of the Uncompressed HD and 4K Videos Using Terahertz Wireless Links

Kathirvel Nallappan, *Graduate Student Member, IEEE*, Hichem Guerboukha, *Graduate Student Member, IEEE,* Chahé Nerguizian, *Member, IEEE*, and Maksim Skorobogatiy, *Senior Member, IEEE*

*Abstract*— Taming the Terahertz waves (100 GHz-10 THz) is considered the next frontier in wireless communications. While components for the ultra-high bandwidth Terahertz wireless communications were in rapid development over the past several years, however, their commercial availability is still lacking. Nevertheless, as we demonstrate in this paper, due to recent advances in the microwave and infrared photonics hardware, it is now possible to assemble high performance hybrid THz communication systems for real-life applications. As an example, in this work, we present design and performance evaluation of the photonics-based Terahertz wireless communication system for the transmission of uncompressed 4K video feed that is built using all commercially available system components. In particular, two independent tunable lasers operating in the infrared C-band are used as a source for generating the Terahertz carrier wave using frequency difference generation in a photomixer. One of the IR laser beams carries the data which is intensity modulated using the $LiNbO_3$ electro-optic modulator. A zero bias Schottky detector is used as the detector and demodulator of the data stream followed by the high-gain and low-noise pre-amplifier. The Terahertz carrier frequency is fixed at 138 GHz and the system is characterized by measuring the bit error rate for the pseudo random bit sequences at 5.5 Gbps. By optimizing the link geometry and decision parameters, an error-free (BER<$10^{-10}$) transmission at a link distance of 1m is achieved. Finally, we detail integration of a professional 4K camera into the THz communication link, and demonstrate live streaming of the uncompressed HD and 4K video followed by analysis of the link quality.

## I. INTRODUCTION

THE internet protocol data traffic is continuing its exponential increase and is expected to reach over 278 Exabytes per month by 2021 [1]. Similarly, the ever-increasing wireless communications data rate in the commercial markets is expected to be 100 Gbps within the next 10 years [2]. To meet the bandwidth demand, a shift towards higher carrier frequencies has been considered as a solution [3-5]. The terahertz (THz) frequency band (Frequency:100 GHz to 10 THz, Wavelength: 3 mm to 30 μm) is seen by many as the next frontier in wireless communications [6, 7]. Most recently, the long distance (>2 Km) wireless links operating in the THz band with a carrier frequency of 120 GHz were reported providing data rates of 10 Gbps and 20 Gbps using amplitude shift keying (ASK) and quadrature phase shift keying (QPSK) respectively [8-11].

At the same time, maturing the THz wireless communication technologies from laboratories into commercial applications is facing multiple challenges. Two major technologies exist in establishing THz wireless communication links: electronics-based frequency multiplier chains and photonics-based frequency difference generation [12]. Electronics-based approaches offer high powers (thus longer link distances), but at lower carrier-wave frequencies (<100 GHz), thus limiting the communication data rates. On the other hand, photonics systems suffer from lower power budgets due to inefficiency in optical to THz conversion, but offer potentially higher data rates at much higher carrier frequencies (>100GHz) [12].

From the prospective of telecommunication applications, one of the key advantages offered by infrared (IR) photonics is its ability to interface directly with the already existing fiber-based network equipment [12, 13]. Therefore, integration of the optics-based THz wireless transmitters with existing IR photonics networks can be done in a seamless fashion. Additionally, high tunability of the THz carrier frequency (between 20 GHz - 3.8 THz [14-16]) is easily achievable using photomixing, thus higher carrier frequencies and, hence, higher data rates are readily achievable in optics-based THz systems. Recent advances in the uni-traveling carrier photodiodes (UTC-PD)–based THz photomixers give a new hope for commercial applications of photonics-based THz communication systems [17-22]. Such devices offer relatively

K. Nallappan is with the Department of Electrical Engineering and Department of Engineering Physics, Polytechnique Montréal, Québec, H3T 1J4 Canada (email: kathirvel.nallappan@polymtl.ca).

H.Guerboukha, and M.Skorobogatiy are with the Department of Engineering Physics, Polytechnique Montréal, Québec,    H3T 1J4 Canada (email: hichem.guerboukha@polymtl.ca & maksim.skorobogatiy@polymtl.ca).

C. Nerguizian is with the Department of Electrical Engineering, Polytechnique Montréal, Québec, H3T 1J4 Canada (email: chahe.nerguizian@polymtl.ca).

This work was supported by the Canada Research Chair I program and the Canada Foundation for Innovations grant (Project No: 34633) in Ubiquitous THz Photonics of Prof. Maksim Skorobogatiy.

high powers (~1 mW) even at higher THz frequencies (~300 GHz) [23, 24]. By using a hybrid approach incorporating both UTC-PD as the emitter and solid-state devices such as Schottky diodes as receivers, several high-speed THz communication links have been demonstrated in recent years. Data rates of 48 Gbps at 300 GHz [25] and 50 Gbps at 330 GHz [26] have been achieved in real-time measurements using amplitude modulation of the optical signals. Similarly, using higher order modulation and offline signal processing techniques, 75 Gbps at 200 GHz [27], 100 Gbps at 237.5 GHz [28], 46 Gbps at 400 GHz [29] and 60 Gbps at 400 GHz [30] have been demonstrated. Using multi-channel modulation and single emitter configuration, a higher data rate of 160 Gbps has been reported recently [31]. As a practical application, a few works have shown successful transmission of HD and 4K video in the lower THz band [10, 32-36]. Similarly, using photonics based THz emitter and heterodyne receiver operating at 600 GHz, a successful transmission of HD video has been demonstrated over a short range (0.5 m) with a received THz power of 10 nW [37]. All these demonstrations have been carried out using proprietary devices, particularly the photomixer and detector electronics.

We now review some of the bandwidth hungry applications that can profit greatly from the THz communication systems. With ever increasing wireless data rate, mobile backhaul for transferring large bandwidth signals between base stations and end users is one of the primary goals using THz band. Employing a highly efficient wireless backhaul service by connecting multiple base stations with small cell size will ensure high data rate to the end users [38]. Thus, enabling photonics based wireless links (millimeter to THz frequency band) offers high efficiency and cost-effective solution to the mobile backhauling and 'last mile' connections [39-41]. Similarly, the THz communication systems can be used to establish short range high-speed wireless links for applications such as chip-to-chip communications, KIOSK downloading, high-speed indoor wireless LAN to name a few [42]. Installation of several high-speed access points in crowded areas such as airports, metro stations, shopping malls could enable the users to receive a high-quality video content, software updates etc. in a short duration.

Among several bandwidth hungry applications, wireless transmission of uncompressed HD, 4K and 8K video finds importance in applications such as education, entertainment, telemedicine, security, video conferencing to name a few. It is reported that the market value of 4K technology will reach $102.1 billion by 2020 [43]. Wireless streaming of high quality videos is already employed in biomechanical analysis, unmanned aerial vehicle (UAV) unmanned ground vehicle (UGV) and telemedicine. In biomechanics, motion analysis (sports person for example) is an important tool for the evaluation of movements that can be also used to improve the performance or the well-being of a person [44]. In this application, the video cameras that capture the motions are portable devices with low storage capacity that use wired connections and off-line information processing. On the other hand, if the uncompressed video content is transmitted wirelessly to a remote workstation, the studies can be performed in real-time. Moreover, with the increasing popularity of 3D cameras for motion studies, the bandwidth requirement for wireless streaming is doubled, thus creating challenges in real-time transmission and recording.

In the medical field, the recent research indicates that the survival rate of the patient during emergency situations depends on the efficiency of pre-hospital care [45, 46]. It also suggests that for the transmission of HD video from the remote location (ambulance for example), to the hospital, a hybrid optical-wireless technology with low latency would be the ideal solution. An unprecedented level of detail to judge the patient condition can be inferred from the video using a zoom-in feature with uncompressed 4K or 8K cameras at the remote pre-medical center [47]. Furthermore, the wireless streaming of uncompressed 4K video is of great importance in surveillance, security and ground mapping applications. An important factor that limits performance of systems employed in these applications is the quality of the received video [48]. By streaming the uncompressed 4K video wirelessly, it is possible to carry out the analysis in real-time with high accuracy. In the broadcasting sector, capturing and telecasting the live events such as sports using high resolution 4K camera is also attracting much attention. For example, NHK (Japan broadcasting corporation) has already started trial experiments by telecasting 8K video using proprietary devices for Olympic games that to be held in 2020 [49]. All of the abovementioned examples suggest that transmission of the uncompressed 4K/8K videos can provide a significant improvement to a variety of important applications.

Although several high-speed communication links have been already demonstrated using proprietary devices, the transfer of such systems to the market is still in the infancy due to lack of many enabling components commercially. Particularly, when implementing optics-based THz communication systems, the components such as electro-optic modulator, coupler, isolators etc. are readily available but the challenges arise in terms of low output power of the THz photomixer, lower sensitivity of fast Schottky detectors, non-availability of wide band interconnects and low noise digital amplifiers. On the other hand, the components for millimeter wave communications, particularly at 60 GHz band is started entering the commercial market. The power amplifiers below 100 GHz already exist delivering the output power in the order of Watts increasing the link distance and channel capacity. While millimeter wave communication is the immediate solution for high speed wireless applications, the offered data rate is still two orders of magnitude below the expected demand [50]. However, to increase the spectral efficiency, a complex hardware architecture need to be implemented.

Recent developments of hardware components in the THz band promises several game-changing applications. Thus, the availability of continuous wave THz photomixers with reasonable output power (~65 µW at 100 GHz-photomixer by Toptica Photonics) enables a variety of applications including high resolution spectroscopy, imaging and communication studies. For THz communication applications, to the best of our knowledge, NTT-Electronics corporation is the only commercial supplier of the waveguide integrated UTC-PD's with the output power of ~200 µW at lower THz frequencies (<200 GHz). Similarly, at the receiver end, the commercial zero bias Schottky diode (ZBD) based detectors offers broad bandwidth but with lower responsivities (~2000 V/W). The availability of low noise digital amplifiers with broad bandwidth (~10 GHz) and high gain (>30 dB) for the amplification of the demodulated baseband signal is still a challenge. However, as we discuss in this paper, it is possible now to establish a short- range THz communications using all commercial components.

In this article, encouraged by the recent advances in the THz and IR photonics, we show the possibility of assembling a high-performance THz communication system for real-world applications, by borrowing off-the-shelf commercial components from various communication technologies. We present the design and the performance evaluation of the photonics based THz communication system that is built using all commercially available system components. As a practical application, we detail the integration of a 4K camera into the THz communication link and demonstrate the live streaming and recording of the uncompressed HD and 4K videos, followed by analysis of the link quality. The paper is organized as follows. Section II presents the experimental setup for the photonics based THz communication system followed by the characterization of the communication system components in Section III. Section IV presents the Bit error rate (BER) measurement results and finally Section V shows the integration and demonstration of wireless transmission of the uncompressed HD and 4K video streams.

## II. TERAHERTZ COMMUNICATION SYSTEM

The schematic of the photonics-based THz wireless communication system is shown in Fig.1. One of the two laser beams is intensity-modulated and then amplified using the Erbium-doped fiber amplifier (EDFA). The laser beams are combined using the 3 dB coupler and then injected into the photomixer for THz generation.

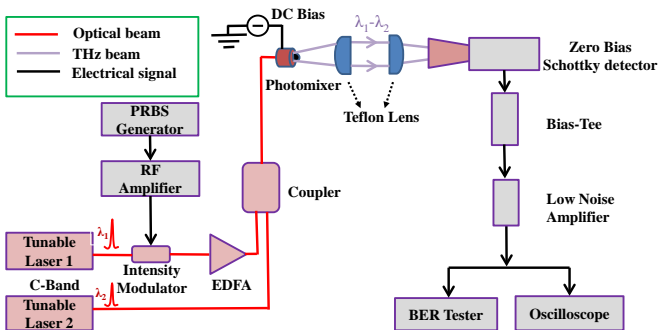

Fig 1: Schematic of the photonics-based THz wireless communication system.

In the receiver side, a zero-bias Schottky diode (ZBD) is used to detect and directly demodulate the incoming THz signal. The DC field present in the demodulated baseband signal is filtered using Bias-Tee and amplified using a low noise amplifier (LNA) for further signal processing. The components of the THz communication system are detailed in what follows.

### A. Laser source

Two tunable distributed feedback (DFB) laser diodes operating in the C-band (Toptica Photonics) with an output power of 30 mW each are used to optically drive the THz emitter. The linewidth of each laser is ~0.5 MHz on a short time scale (100 µs), while the laser frequency may drift over a few tens of MHz in a few minutes depending on the stability of the ambient conditions. Both diodes are mounted in a compact laser head and the output beams are delivered via a polarization-maintaining (PM) single-mode fiber. The wavelength tuning behavior of the lasers is calibrated [51] and the output THz frequency -i.e., the frequency difference of both lasers- is tuned using an Field Programmable Gate Array (FPGA)-based control module ("Teracontrol" by Toptica Photonics).

### B. Data Modulation

One of the lasers is intensity-modulated (On-Off keying modulation, OOK) using an external electro-optic Mach-Zehnder modulator (Thorlabs-LN81S-FC). The operating bias point of the modulator is controlled by the modulator driver unit (Thorlabs-MX10A) which also acts as an RF amplifier. The modulator is locked to the quadrature bias point with positive slope which is monitored continuously using a dither tone. The frequency and amplitude of the dither tone is set to 3KHz and 600 $mV_{pp}$ respectively. The THz communication system is characterized by transmitting the Non-Return to Zero (NRZ) pseudo random bit sequence (PRBS) data which is generated by the pulse pattern generator (PPG) unit integrated in the test equipment (Anritsu-MP2100B). The generated PRBS data with a peak-to-peak amplitude of 0.4 $V_{pp}$ is connected to the modulator driver unit, while the amplified electrical data with amplitude of 5 $V_{pp}$ is fed to the optical modulator. The optical output power after the electro-optic modulator is ~200 µW and is fixed to a constant of 110 µW using the variable optical attenuator to avoid any power fluctuations during the measurements. It is further amplified using EDFA amplifier (Calmar laser-AMP-PM-18) to ~13 mW. The pump current of the EDFA is adjusted to have an average output power similar to the unmodulated direct laser beam.

### C. THz generation

A 3 dB fiber coupler is used to combine the direct and modulated laser beams and the output is fed to the photomixer (Toptica Photonics) for THz generation. The photomixer features a silicon lens (6.05 mm ± 0.1 mm thickness and 10.0 ± 0.1 mm diameter) to pre-collimate the output THz beam. The output THz power of a photomixer is shown in Fig.2, which is measured using calibrated Golay cell [52]. We see that the output power is not uniform, and the high power is obtained mostly below 200 GHz. Two plano-convex lenses (Thorlabs-LAT100) with a focal length of 100 mm and a diameter of 50 mm are used to collimate the THz beam in free space and focus it again into the detector.

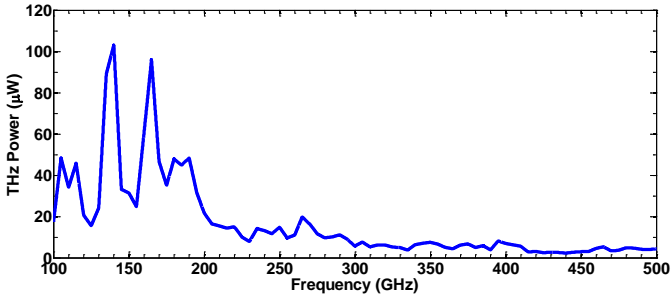

Fig 2: THz output power from the photomixer measured using a calibrated Golay cell with 5 GHz spectral resolution (Courtesy of Toptica Photonics).

*D. THz detection and demodulation*

A ZBD with a horn antenna is used to detect and demodulate the incoming THz signal. The choice for the THz carrier frequency is determined by the product of the output power of the THz emitter and the responsivity of the ZBD. Particularly, two ZBDs (Virginia diodes WR8.0ZBD-F and WR6.5ZBD-F) with a working range below 200 GHz were used in the analysis. The responsivity data for both detectors is shown in Fig.3.

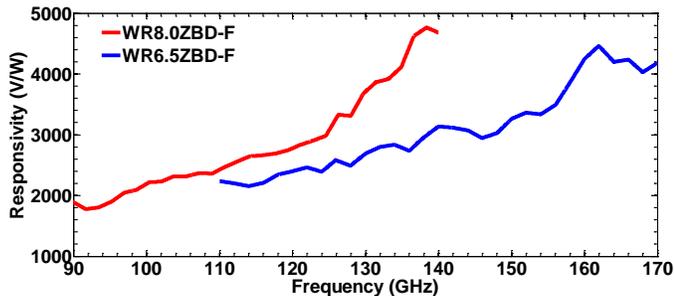

Fig.3. Responsivities of commercial ZBDs WR8.0ZBD-F and WR6.5ZBD-F (Courtesy of Virginia Diodes).

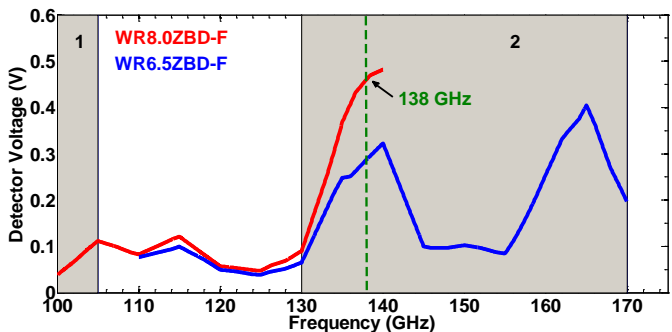

Fig.4. Estimate of the maximal voltage at the ZBD's output when used with Toptica THz photomixer. Grey areas correspond to the THz atmospheric transmission windows as identified in [53].

The maximal voltages at the output of the two ZBDs are estimated by multiplying the emitted THz power as presented in Fig. 2 by the responsivity of a corresponding ZBD presented in Fig. 3. In Fig. 4, we plot such voltages as a function of the THz frequency and conclude that WR8.0ZBD-F ZBD with the conical horn antenna (Virginia diodes WR8.0CH) operating at 138 GHz results in the maximal detected signal amplitude for a back-to-back emitter/detector arrangement. This conclusion was also confirmed experimentally by performing a frequency sweep and detecting the frequency of maximal detected signal at a fixed short link length of ~30 cm. Furthermore, this frequency lies in the second atmospheric transmission window above 100 GHz [53] (grey area in Fig. 4), making it a natural choice as a carrier frequency for THz communications.

Finally, after ZBD, a Bias-Tee (Minicircuits-ZFBT-6GW+) is connected in sequence to block the DC voltage and let only the AC signal from 100MHz to 6 GHz passing to the next stage. Followed the Bias-Tee, a high gain (> 32 dB) and low-noise amplifier (Fairview Microwave-SLNA-030-32-30-SMA) is used to amplify the demodulated baseband signal. The noise figure and the bandwidth of the low noise amplifier (LNA) are 2.5 dB and 3 GHz respectively.

### III. CHARACTERIZATION OF THE COMPONENTS USED IN THE COMMUNICATION SYSTEM

A single-ended, NRZ, PRBS data with a bit rate of 5.5 Gbps and a pattern length of $2^{31}-1$ is used as the test signal to characterize the THz communication system. The peak-to-peak amplitude of the test signal is set to 0.4 $V_{pp}$. The optical spectra of the DFB lasers (before and after modulation) with a frequency separation of 138 GHz are measured using the Optical spectrum analyzer (Anritsu-MS9740A) with a spectral resolution of 0.03 nm as shown in Fig. 5. As discussed in the previous section, one of the lasers is intensity modulated and amplified using the EDFA. First, the PRBS data is turned OFF and the optical output power from the modulator is measured. We observe a high-power loss (>10 dB) due to the optical modulator. The output from the optical modulator is connected again to the modulator driver unit, where the adjustments of the final output power can be done using variable optical attenuator based on user requirement. It is possible to choose either constant output power or constant attenuation mode and we used constant output power mode to avoid any fluctuations in the optical power. Since the optical power from the modulator is low (~200 µW), we fixed the variable optical attenuator to provide the output of 110 µW to further reduce the risk of power fluctuations. Next, an EDFA is used and by adjusting its pump current, the amplitude of the laser beam is kept similar to that of the direct laser beam (blue curve in Fig.5 pointing unmodulated + amplified laser signal).

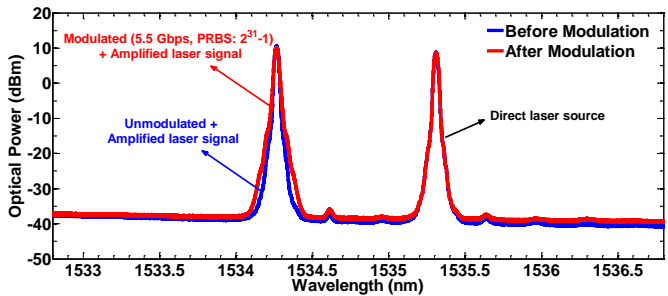

Fig.5. Spectra of the input optical signal injected into the THz photomixer. The SNR of >40 dB is achieved.

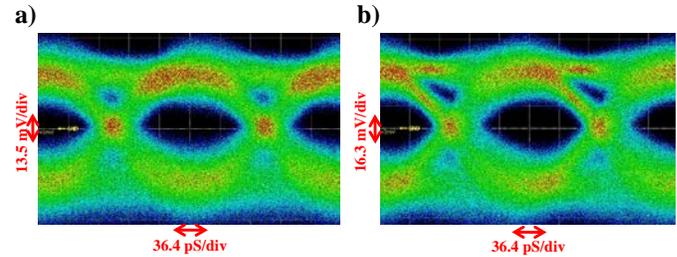

Fig.6. Eye patterns for the test signal (PRBS: 5.5 Gbps, $2^{31}$-1), a) without Bias-Tee and b) with Bias-Tee.

When the PRBS data is turned ON, the laser beam is now modulated, and we observe an increase in the bandwidth of the modulated laser signal as well as the optical noise floor (~1 dB) when compared with the unmodulated laser beam as shown in Fig.5 (red curve pointing modulated + amplified laser signal). However, a signal-to-noise ratio (SNR) greater than 40 dB is achieved after the modulation. The total optical input power (combined laser power) injected into the THz photomixer is 26 mW.

At the receiver side, the Bias-Tee (operational bandwidth 100 MHz - 6 GHz) acts as a band-pass filter as well as to block the DC voltage passing to the next stage (LNA) and thereby to increase the performance. Although the LNA is AC coupled, we observe the improved performance while using the Bias-Tee. The circuit of the Bias-Tee consists of a capacitor and inductor connected in parallel with three external ports (RF+DC-input, DC and RF-output). The DC port is connected to a 50Ω load such that the floating baseband voltage signals is adjusted to a common ground and thereby improves the performance with higher eye amplitude when compared with the measurement without Bias-Tee.

To ascertain the impact of the Bias-Tee on the signal quality, an eye pattern is recorded after the LNA using a high-speed oscilloscope (Anritsu-MP2100B) without and with Bias-Tee as shown in Fig.6. The link distance and the DC bias voltage of the photomixer are fixed at 1 m and -1.9 V respectively for this measurement. The eye pattern measured using the Bias-Tee shows higher eye (opening) amplitude (~18% increase) when compared to the eye pattern taken without Bias-Tee While it improves the performance of the communication system by blocking the DC voltage, at the same time the Bias-Tee might cause some problems when transmitting almost constant bit patterns (like long continuous string of ones or zeros). This is related to the fact that the low frequency components present in the long continuous pulse are filtered out by the Bias-Tee. For example, the lowest non-zero frequency component present in the PRBS test signal with a bit rate of 5.5 Gbps and a pattern length of $2^{31}$-1 is 2.56 Hz ($5.5 \cdot 10^9 Hz/2^{31}$) and it is filtered out by the Bias-Tee along with its harmonics up to 100 MHz, affecting the quality in the signal reconstruction. While keeping in mind potential issues with long constant bit patterns, we nevertheless observe that using Bias-Tee generally leads to higher performance in our experiments.

Because of the lower responsivity values of ZBD's (~2000 V/W) in the THz region, the corresponding output voltage from them after THz detection and baseband signal demodulation is also lower depending on the received THz power. To amplify the baseband signal, a digital amplifier with broad bandwidth (~10 GHz), low noise-figure (<3 dB), high gain (>30 dB) and SMA connectors (since ZBD's are terminated with SMA connectors) is preferred. To the best of our knowledge, such amplifiers satisfying all the above requirements are not available in the commercial market. Therefore, we used a LNA that is designed for analogue signals in our communication system. However, it can still be used for digital signals.

To test the use of analogue LNA with digital signals, the peak-to-peak amplitude of the test signal is attenuated using a fixed attenuator to ~30 mV$_{pp}$ in order to protect the test equipment from the damage after amplification (damage threshold of electrical oscilloscope is ±2V). The evaluation of the waveform such as jitter and eye crossing are automated in the test equipment. The RMS jitter of the digital test signal after attenuation (but before LNA) was measured to be 1.28 pS and 1.48 pS for 3 Gbps and 5.5 Gbps respectively. The attenuated digital test signal is then given as the input to the analogue LNA. The measured eye pattern from the output of the LNA is shown in Fig.7. The RMS jitter measured after the LNA is 4.23 pS and 6.74 pS for 3 Gbps and 5.5 Gbps respectively. Similarly, the eye crossing percentage of the digital signal after the LNA deviates from ideal 50% to ~48% and ~40% for 3 Gbps and 5.5 Gbps respectively after amplification. This may be due to the fact that the duration of digital 0 is longer than digital 1 causing the reduction in the eye crossing percentage from the ideal value. Therefore, we conclude that while analogue LNA can be used for amplification of the digital signals, it can also significantly increase the timing jitter and deviation in the eye crossing percentage in the system while increasing the bit rate.

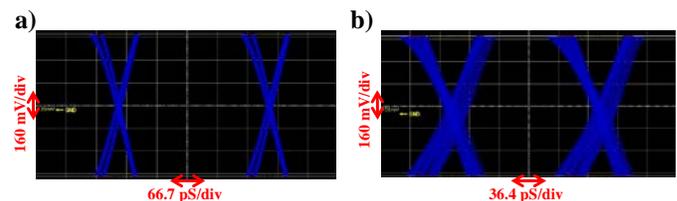

Fig.7. Eye patterns from the output of the analogue LNA using digital a) 3 Gbps and b) 5.5 Gbps signals.

## IV. EVALUATION OF THE THZ COMMUNICATION SYSTEM USING BER MEASUREMENTS

The performance of the THz communication system was then characterized using Bit Error Rate (BER) tester (Anritsu-MP2100B). The BER was first characterized as a function of the photomixer bias voltage, which is a key parameter that defines the emitted THz power. To record a highly consistent BER data in short measurement duration, we choose the target BER of $10^{-12}$ in our experiments. For the target BER of $10^{-12}$, and the 5.5 Gbps bit rate, the duration of a single measurement is calculated as measurement time = 1/(target BER · bit rate) ≈ 182 sec. By fixing the link distance to 100 cm, and the decision threshold to 0 mV, the BER was measured after the LNA by varying the DC bias voltage of the photomixer (Fig. 8).

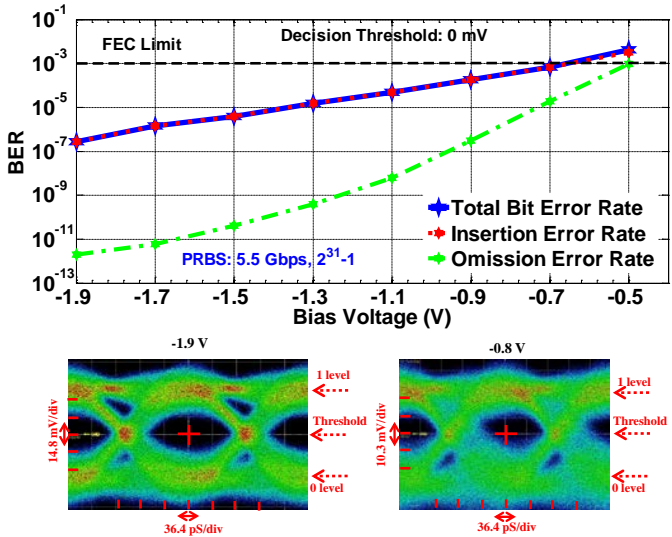

Fig.8. Measured BER as a function of the photomixer bias voltage for the PRBS test signal with 5.5 Gbps bit rate and $2^{31}$-1 pattern length.

The damage threshold limit of the DC bias voltage for the THz photomixer is -2 V and therefore, we limited our measurements to a maximum bias of -1.9 V. The lowest BER of ~$10^{-7}$ was achieved for the bias voltage of -1.9 V. By gradually decreasing the bias voltage, it was found that the BER increased exponentially fast. At the limit of the Forward Error Correction (BER ~$10^{-3}$), the bias voltage was -0.7 V. From Fig. 8 we also observe that insertion errors (digital 0 is mistaken for the digital 1) contributes more to the total BER when compared to the omission errors (digital 1 is mistaken for the digital 0). This is due to the vertical asymmetry of the eye pattern in the received signal. As seen from the eye patterns (inserts in Fig. 8), the digital 0 is closer to the decision threshold than digital 1, which gives rise to the higher insertion error rate.

Next, we have characterized BER as a function of several communication link parameters such as link distance, as well as angular deviation of the detector antenna from the principal direction of signal propagation. Thus, in the first experiment, we have fixed the link distance of 100 cm and the photomixer DC bias voltage to -1.9 V. The BER was then measured by varying the alignment angle of the ZBD detector antenna with respect to the signal propagation direction as defined by the emitter antenna (see the insert in Fig.9). Experimentally this was accomplished by changing the angle of rotation using high precision rotation mount (Thorlabs-PR01), where the detector antenna is fixed on top of it. The total BER was found mainly due to the insertion errors (see Fig. 9) due to asymmetry of the eye pattern in the vertical direction. From these measurements, we found that in order to stay within the FEC limit, the detector has to be within ~4.8 degrees of the principal beam propagation direction.

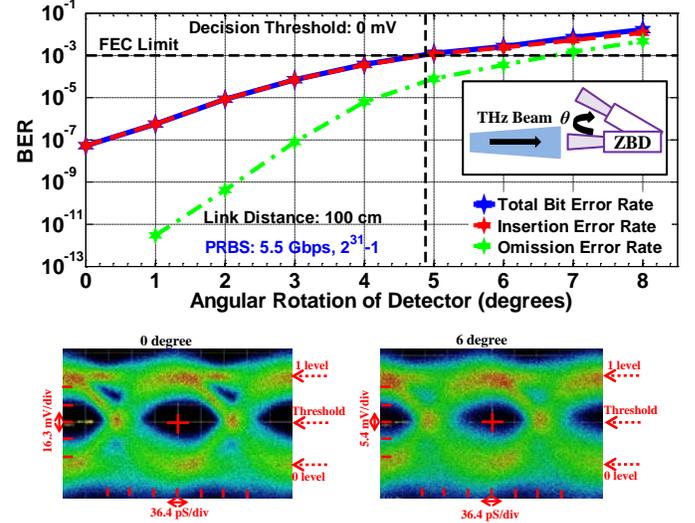

Fig.9. Measured BER as a function of the angle $\theta$ between the detector and the signal propagation direction for the PRBS test signal with 5.5 Gbps bit rate and $2^{31}$-1 pattern length. Insert: schematic of the experiment.

In the second experiment, the BER was measured as a function of the link distance. First, we fixed the decision threshold to 0 mV and measured the BER at several link distances. As our system was mounted and aligned on the optical table, the measurements were limited to 175 cm long links. The measured BER is shown in Fig. 10 (a), we note that even at 175 cm link length, the total error is below the FEC limit. Due to vertical asymmetry in the eye diagram, the insertion error is again much higher than the omission error. This asymmetry can be addressed by adjusting the value of the decision threshold in order to equalize the insertion and omission errors, which also leads to the lower total BER.

Therefore, in the follow-up experiment we optimized the system performance by finding the optimal value for the decision threshold at each link length in order to equalize the insertion and omission errors. In Fig. 10 (b) we plot performance of the optimized THz communication system as a function of the link distance, where omission and insertion errors are balanced. For comparison, in the same figure we present total error of the unbalanced system (same as Fig. 10(a)) and note that decision threshold optimization indeed reduced the total BER, which is especially pronounced at shorter link lengths.

In passing we note that for shorter communication links, the main reason for the BER increase with the link distance is divergence of the THz beam, and a consequent reduction of the THz power at the detector site. Particularly, due to diffraction on the collimating lens, the diameter $D_{THz}$ of the THz beam as a function of the link distance $L$ can be approximated as $D_{THz} \sim L \cdot \lambda_c / D_l$, where $\lambda_c$ is the wavelength of THz carrier wave, and $D_l$ is the lens diameter. If two identical lenses are used to collimate the THz beam and focus that beam onto a detector, then with $P_0$ power emitted by the photomixer, only $P = P_0(D_l/D_{THz})^2 \sim P_0(D_l^2/(L \cdot \lambda_c))^2$ will arrive to the detector. From this we conclude that two strategies can be pursued in order to increase power budget in case of short communication links. These include either using collimating lenses of larger diameter $D_l$, or using higher THz carrier frequencies (lower $\lambda_c$). For longer communication links, THz absorption in atmosphere becomes important, which further decreases the power budget and increases detection errors.

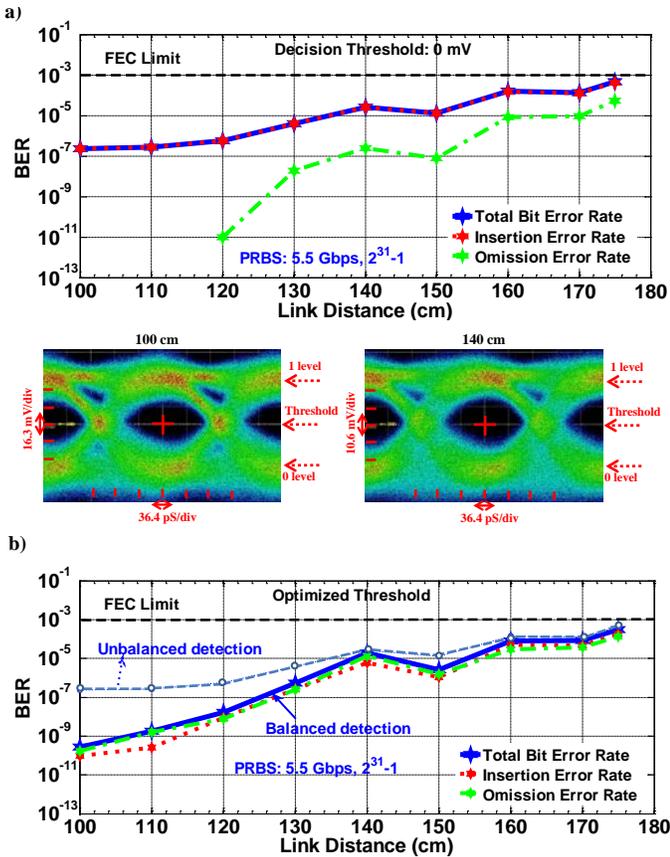

Fig. 10. Measured BER as a function of the communication link length for the PRBS test signal (5.5 Gbps, $2^{31}$-1). (a) Unbalanced system, the decision threshold is fixed at 0 mV. (b) Balanced system: for each link length, the decision threshold is optimized to balance the insertion and omission rates.

## V. TRANSMISSION OF UNCOMPRESSED HD AND 4K VIDEO USING THZ COMMUNICATIONS

As indicated in the introduction, one of the important practical applications of THz communications is wireless transmission of high-quality video. In what follows, we first detail integration of the professional 4K camera into the THz communication system detailed earlier, and then demonstrate successful transmission of the uncompressed HD and 4K videos at 60 frames per second (fps) and 30 fps respectively. Additionally, we discuss two different strategies for the integration of 4K cameras at the transmitter side of a THz communication system depending whether electrical or optical outputs of the camera are used.

A 4K camera (Blackmagic studio camera) with two output ports (optical and electrical) was used in our experiments. A standard output from a 4K camera is an uncompressed video with 10-bit color depth and Chroma sub-sampling of 4:2:2. The peak-to-peak output voltage from the SDI (Serial Digital Interface) electrical cable is 800 mV$_{pp}$. In Fig.11 (a) we present integration of a 4K camera with THz transmitter when using electrical output from the camera. The electrical cable from the 4K camera has a BNC connector with an impedance of 75Ω, whereas the RF amplifier present in the optical modulator driver has a SMA connector with an impedance of 50Ω. Therefore, an impedance matching pad with a bandwidth of 2 GHz (Fairview Microwave-SI 1560) is used in the design for the demonstration of 4K video transmission. To our knowledge, currently, this is the only commercially available impedance matching pad with BNC to SMA connectors that offers broadband operation. However, due to 2 GHz bandwidth limitation of the impedance matching circuit, we find that transmission of the uncompressed 4K video is limited to 30 fps (~6 Gbps). While feeding the video signal from the camera to the optical modulator, the control settings (amplified RF voltage level and modulator bias voltage) for RF amplifier and the external modulator are kept similar to the BER measurements.

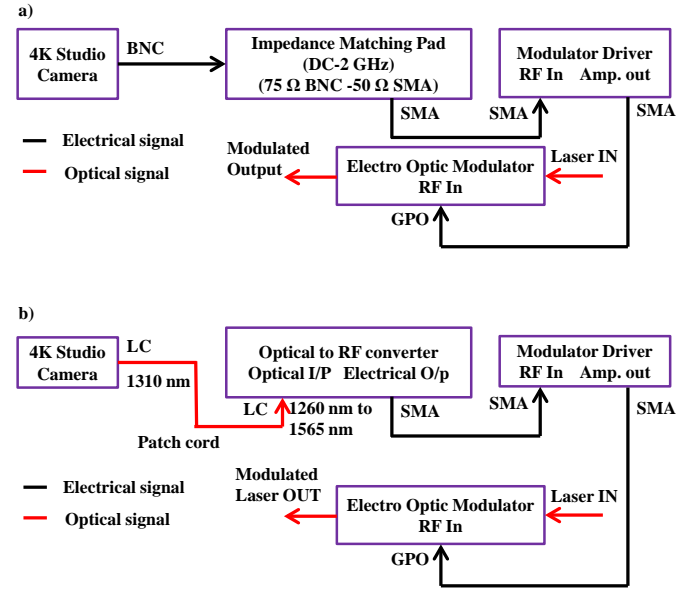

Fig.11. Block diagram for the integration of a 4K video camera with transmitter of the THz communication system using a) electrical and b) optical signal output from the camera.

Alternatively, in Fig.11 (b) we present integration of a 4K camera with THz transmitter when using optical signal from the camera output. In order to use optical video signal with the THz transmitter, it has to be converted to electrical signal to feed the optical modulator. Therefore, an optical transceiver (Finisar-FTLX1471D3BCL) which is integrated with the test equipment (Anritsu-Mp2100B) is used for the optical to electrical conversion. The optical signal from the 4K camera was carried via a single mode fiber. The electrical output of the optical transceiver features SMA connector with 50Ω impedance, which was then connected to the SMA input RF port of the optical modulator driver. Using both designs presented in Fig.11, an uncompressed 4K video up to 30 fps can be transmitted. The advantage of using electrical output from the camera (Fig.10 (a)) is that, the impedance matching pad is compact, less expensive and does not require the power supply for the operation. However, the RF cable is bulky to handle and suffers from high signal loss as the length increases. Conversely, while using optical output from the camera (Fig.11 (b)), we benefit from low losses of the optical fiber and it also supports large bandwidth of the baseband video signal. At the same time, the maximum data rate of the optical transmitter supported by the 4K camera is 6 Gbps, limiting the uncompressed 4K video frame rate to 30 fps. For our demonstration, we used the optical signal from the 4K camera to integrate with the THz transmitter (Fig.11, b) as it favors easy handling and integration using plug-in connectors.

While so far, we have discussed transmitter side of the THz communication system for video transmission, we now focus on the receiver side. At the receiver side, in order to present and analyze the transmitted video, we have explored two options including a 4K-capable display (ASUS-MG28UQ), and a 4K-capable video receiver (Blackmagic) and a storage unit. The received, demodulated and amplified electrical signal is converted into optical signal using optical transceiver of the Anritsu-MP2100B test equipment for further processing as shown in Fig.12. It is possible to use the impedance matching pad (similar to the transmitter integration) after the LNA without converting it to the optical signal to connect it to the computer interface card. But, the high frequency components (>2 GHz) present in the baseband signal will be filtered by the impedance matching pad (Bandwidth: DC-2 GHz), which is not favorable in signal reconstruction. Therefore, we preferred to use the optical conversion as the fiber can carry all the frequency components present in the amplified baseband signal, favoring efficient signal reconstruction and easy handling. Particularly, the electrical to optical conversion corrects the amplitude distortions that are present in the electrical signals (see the eye pattern in Fig.13).

The receiver configuration presented in Fig.12 can be modified depending on the end requirement (display or record). If displaying the 4K video is the only requirement, an optical to HDMI converter (Blackmagic-Teranex Mini Optical Fiber) is used. A high-speed computer interface card (Blackmagic-Decklink 4K Extreme 12G) is employed when both recording and displaying the 4K video simultaneously is required. The computer interface card supports SDI (electrical), HDMI and optical input/output interfaces allowing to record the received uncompressed video signal on the solid state hard drive (SSD). In order to test the capability of the SSD to record uncompressed video signals, we first analyzed the writing and reading speed of the SSD with PCIe 3.0 interface (SAMSUNG-960 PRO M.2 2 TB NVMe PCIe 3.0) and SSD with SATA interface (SAMSUNG-MZ75E500) using Blackmagic disk speed test software. Particularly, we connected four SSD's (500 GB each) with SATA interface using RAID 0 configuration to increase the data transfer speed. The analysis showed that it is possible to write and read 2000 MB/s using SSD with PCIe3.0 interface whereas, the data transfer speed of the SSD with SATA interface is slower (Writing speed: 1197.4 MB/s and Reading: 1473.2 MB/s). Therefore, we used the SSD with PCIe3.0 interface to record the uncompressed HD and 4K video.

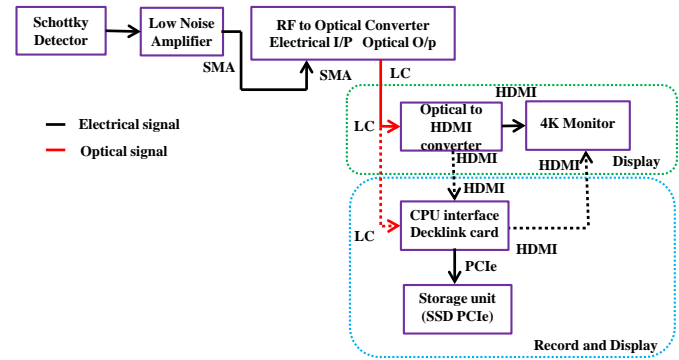

Fig.12. Block diagram for the receiving, visualization and recording of the transmitted uncompressed HD and 4K video.

Before performing the video transmission experiment, the BER measurement is conducted again using PRBS data in order to compare whether the total BER after electrical to optical conversion is similar to the total BER measured before conversion. Since, the BER can be measured only for the electrical signals in our test equipment, the optical signal is further converted to electrical signal using optical transceiver (only for BER measurements in this case). The BER measurement is conducted at several link distances with fixed decision threshold (0 mV) as shown in Fig.13. However, the eye pattern in the inset of Fig.13 is measured for optical signal. We observe that the total BER after optical conversion is slightly increased when compared to the BER that is measured with the electrical signal after LNA (Fig.10 (a)). Therefore, we conclude that the effect of converting the electrical signal to optical signal for the video transmission experiment is negligible.

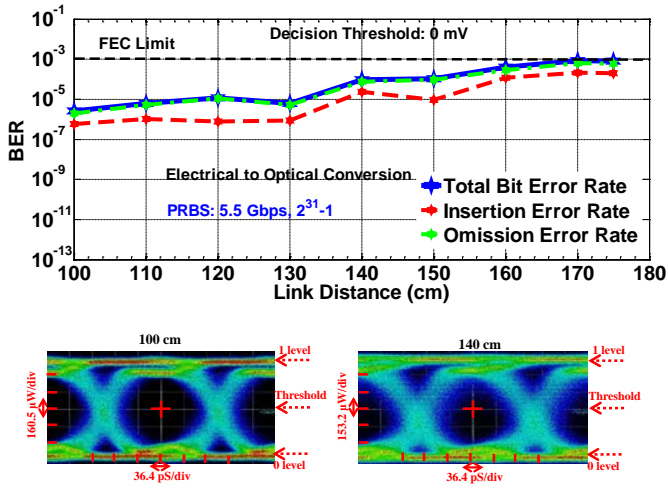

Fig.13. BER versus communication link distance after electrical to optical conversion (Fixed decision Threshold: 0 mV)

For the demonstration of video transmission wirelessly, an uncompressed HD (1920 x 1080 at 60 fps) and 4K (3840 x 2160 at 30 fps) videos are transmitted individually through the THz communication system at the carrier frequency of 138 GHz. The corresponding data rates are 2.97 Gbps and 5.94 Gbps for HD and 4K video respectively. For the ease of analysis, the received video is recorded as image data (Digital Picture Exchange format) where individual frame from the video is recorded as an image. The received frames are recorded at a given link distance for a duration of 30 minutes. The number of black frames are identified by taking the small portion (5x5 pixels) in each frame and analyzing the average RGB value. The RGB value of the black frame is zero and therefore any frame that is having the average RGB value above this threshold is identified as a valid frame. The black frame is a tool to measure the performance of the wireless streaming which is caused due to the error in the synchronization packets. The timing synchronization of the digital video is provided by the End of Active Video (EAV) and Start of Active Video (SAV) sequences with a unique hexadecimal word pattern. The hexadecimal word pattern refers to 3FF (all bits are 1), 000 (all bits are 0), 000 (all bits are 0), and XYZ (10-bit word). The XYZ of a 10-bit word corresponds to bit number 8,7 and 6 which is used to indicate whether the video scanning is progressive or interlaced, active video or blanking interval and EAV or SAV respectively. Therefore, any error in this synchronization packet leads to the black frame.

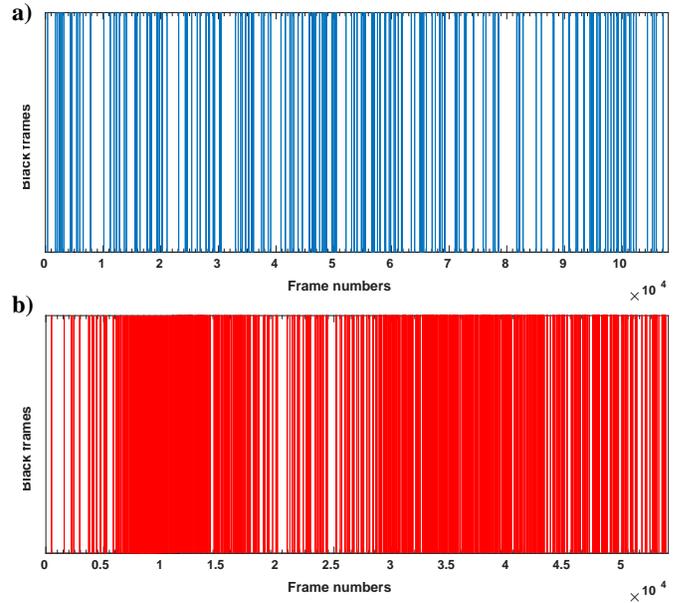

Fig.14. Identified black frames which is recorded for 30 minutes at a link distance of 30 cm. a) HD video (60 fps) and b) 4K video (30 fps)

The identified black frames for the HD (60 fps) and 4K (30 fps) video which is recorded at the link distance of 30 cm for the duration of 30 minutes (HD-108000 frames and 4K-54000 frames) is shown in Fig.14. We see that the possibility of black frame is higher for 4K video transmission when compared with the HD video due to minor instability of the THz communication system during the measurement duration.

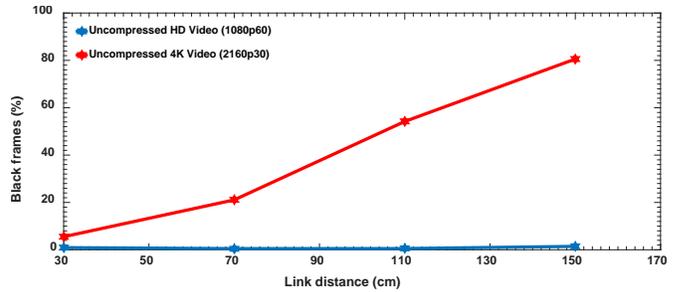

Fig.15. Percentage of black frames for HD and 4K video as a function of link distance.

We also did the similar experiment by increasing the wireless link distance and the percentage of black frames is calculated for both HD and 4K videos as shown in Fig.15. We see that, the percentage of black frames is less than 0.5% for HD video and ~5% for 4K video at the link distance of 30 cm. By increasing the link distance, the percentage of black frames is almost constant for HD video but increases for 4K video. It indicates that the probability of errors in the timing synchronization packets is higher for the 4K video due to high bit error rate.

## VI. CONCLUSION

To conclude, we showed the design and evaluated the performance of a photonics based THz wireless communication system that is built using all commercially available system components. Two independent tunable lasers operating in the infrared C-band are used as the source for the generation of THz carrier wave in the photomixer. A zero bias Schottky diode is used as the detector and demodulator followed by a high gain and low noise amplifier. The optimum carrier frequency (138 GHz) is chosen by analyzing the THz output power and the responsivity of commercially available ZBDs. The performance of the built system is evaluated by measuring the BER for the PRBS data at the bit rate of 5.5 Gbps. By optimizing the decision threshold, an error-free data transmission is achieved at a link distance of 1 m. Finally, we detailed the integration of a 4K camera at the THz transmitter and the video receiver electronics at the THz receiver. The practical application of the built THz system is demonstrated by the successful transmission of uncompressed HD and 4K video. The link quality for the video transmission is analyzed and the percentage of black frames are measured. It is observed that the percentage of black frames is higher for the transmitted uncompressed 4K video due to the high bit errors at the increased link distance, whereas ~99% of the uncompressed HD video is received successfully. The obtained results confirm that it is now possible to realize a short-range THz wireless communication system for commercial applications.

ACKNOWLEDGMENT

We thank Dr. Anselm Deninger (Toptica Photonics), Mr. James Morgente (Anritsu), Mr. Jules Gauthier (Polytechnique Montreal) and Blackmagic support team for their valuable technical discussions. Also, we thank our technicians Mr. Francis Boutet, Mr. Jean-Paul Lévesque and Mr. Yves Leblanc for their assistance.